\documentclass[]{elsarticle} 
\usepackage[hyphens]{url}
\usepackage{lineno} 
\providecommand{\tightlist}{%
  \setlength{\itemsep}{0pt}\setlength{\parskip}{0pt}}

\biboptions{sort&compress} 
\usepackage{graphicx}
\usepackage{booktabs} 

\usepackage[T1]{fontenc}
\usepackage{lmodern}
\usepackage{amssymb,amsmath}
\usepackage{ifxetex,ifluatex}
\usepackage{fixltx2e} 
\IfFileExists{upquote.sty}{\usepackage{upquote}}{}
\ifnum 0\ifxetex 1\fi\ifluatex 1\fi=0 
  \usepackage[utf8]{inputenc}
\else 
  \usepackage{fontspec}
  \ifxetex
    \usepackage{xltxtra,xunicode}
  \fi
  \defaultfontfeatures{Mapping=tex-text,Scale=MatchLowercase}
  
\fi
\IfFileExists{microtype.sty}{\usepackage{microtype}}{}
\ifxetex
  \usepackage[setpagesize=false, 
              unicode=false, 
              xetex]{hyperref}
\else
  \usepackage[unicode=true]{hyperref}
\fi
\hypersetup{breaklinks=true,
            bookmarks=true,
            pdfauthor={},
            pdftitle={Rising Novelties on Evolving Networks: Recent Behavior Dominant and Non-Dominant Model},
            colorlinks=true,
            urlcolor=blue,
            linkcolor=magenta,
            pdfborder={0 0 0}}
\usepackage{setspace}
\usepackage{graphicx}
\DeclareGraphicsExtensions{.pdf,.eps,.png,.jpg}
\usepackage{setspace}
\usepackage{graphicx}
\DeclareGraphicsExtensions{.pdf,.eps,.png,.jpg}

\usepackage[nomarkers]{endfloat}

\begin{document}
\begin{frontmatter}

  \title{Rising Novelties on Evolving Networks: Recent Behavior Dominant and
Non-Dominant Model}
    \author[Big Data Research Center UESTC,Big Data mining and application CIGIT CAS]{Khushnood Abbas\corref{c1}}
   \ead{abbas@cigit.ac.cn}
   \cortext[c1]{Corresponding Author}
      \address[Big Data Research Center UESTC]{Big Data Research Center, University of Electronic Science and
Technology of China, Chengdu 611731, PR China}
    \address[Big Data mining and application CIGIT CAS]{Big Data mining and application center, Chongqing Institute of green and
Intelligent Technology of China, Chinese Academy of Sciences, Chongqing,
401120, P.R. China.}

  \begin{abstract}
  Novelty attracts attention like popularity. Hence predicting novelty is
  as important as popularity. Novelty is the side effect of competition
  and aging in evolving systems. Recent behavior or recent link gain in
  networks plays an important role in emergence or trend. We exploited
  this wisdom and came up with two models considering different scenarios
  and systems. Where recent behavior dominates over total behavior (total
  link gain) in the first one, and recent behavior is as important as
  total behavior for future link gain in second one. It suppose that
  random walker walks on a network and can jump to any node, the
  probablity of jumping or making connection to other node is based on
  which node is recently more active or receiving more links. In our
  assumption random walker can also jump to node which is already popular
  but recently not popular. We are able to predict rising novelties or
  popular nodes which is generally suppressed under preferential
  attachment effect. To show performance of our model we have conducted
  experiments on four real data sets namely, MovieLens, Netflix, Facebook
  and Arxiv High Energy Physics paper citation. For testing our model we
  used four information retrieval indices namely Precision, Novelty, Area
  Under Receiving Operating Characteristic(AUC) and Kendal's rank
  correlation coefficient. We have used four benchmark models for
  validating our proposed models. Although our model doesn't perform
  better in all the cases but, it has theoretical significance in working
  better for recent behavior dominant systems.
  \end{abstract}

 \end{frontmatter}

\section{Introduction}\label{introduction}

Every time we use the Internet (particularly social media platforms,
search engines, email servers) we leave our traces behind which
generates a huge volume of data (Hommes 2002), (Bouchaud, Matacz, and
Potters 2001), (Haldane and May 2011). This huge data carrying a lot of
opportunity for understanding the individual as well as collective
behaviours. Collective behaviours are more important in many domain such
as finance, viral marketing, epidemic disease prevention and so on, by
targeting or controlling individuals or communities when predicting
their behaviour and / or events and trends in societies. For example
from Google trend some researchers infer the real world phenomena such
as geographical location of disease spread (Ginsberg et al. 2008), stock
market prediction (Preis, Moat, and Stanley 2013), which product will be
in demand or who is going to win this election.Likewise, social media
platforms such as Twitter has been used to track human mobility, improve
the response to disasters, etc. Therefore we can say this big data is
carrying ample amount of opportunities for scientists and researchers
from all disciplines to solve this complex world problem (Axtell 2001),
(King 2011), (Vespignani 2009), (Perc 2012). Most of the big data of the
interactions of people in the Internet can be modeled The following two
main different types of networks can be shaped from such data: a)
monopartite, in which the nodes have similar types (i.e., only people or
items) and links reflect a relationship among them; and b) bipartite, in
which the nodes can be from different types (i.e., people and items) and
links show a relationship between different types of nodes. By generally
simply the latter type can be transformed to two different monopartite
networks. Nevertheless, one of the main attributes in such network is
the link formation among the nodes over time which can help to manage
the dynamics of the network and make appropriate decision (Abbasi 2016).

Although the quality item or fit node always show some common
characteristics i.e they compete to get attention. It may be possible in
presence of popular nodes its strength forgotten. Researchers have also
found that in network such as paper citation it may not get attention
for long time (Ke et al. 2015), some gain popularity many times (Cheng
et al. 2016). In all the cases we think as also been proven that recent
activity of the node is one of the important feature of node for future
activity (Gleeson et al. 2014), (ZENG et al. 2013). Popularity or future
link gain prediction is a complex task depending on different factors
like quality or nodes fitness and so on. Content popularity may
fluctuate with time (Eisler, Bartos, and Kert\a'esz 2008), increase over
time or be limited within communities. Most recently researchers have
found that the popularity of online contents like news, blog posts,
videos, mobile app download (Gleeson et al. 2014) in online discussion
forums and product reviews exhibits temporal dynamics.

Emergence or first time occurence is novelty for those who didn't face
it before, which can be innovation also. Innovation is special case of
novelty in a sense it is entertained by collective attention, while
novelty may not. The emergence, novelty , trend or future occurrence
prediction models can be found since quite long time ago such as
(Zanette and Montemurro 2005), (Hoppe 1984), (Simon 1955), (Tria et al.
2014). Emergence prediction theory applies in many areas of life such as
economics (Simon 1955), biology (E. and Yule 1925), physics (Redner
1998) etc. (Simon 1955) has given stochastic model which fits to skewed
or power law distribution or fit to rich-gets-richer phenomenon. It is
obvious that popularity of any item on social media doesn't last
forever. In addition decay rate may vary from item to item. Every type
of node have its own decay rate like (Parolo et al. 2015) have found
research article citation rate decays after some time. Cumulative
advantage is a well known phenomenon seen almost in every evolving
networks, which states that popularity increases cumulatively; the rate
of new link (Either item receives rating in case of Movielens, or a
friend like or comments in case of Facebook wall post activity)
formation for any node is proportional to the observed number of links
which node has received in past. Authors in (D. Wang, Song, and Barabasi
2013) for quantifying long term prediction considered fitness of paper
for future citation gain of the paper. The nature of paper citation
distribution is also found fat tailed or skewed which tells only few
papers carrying most of the citation while most of the papers no
citation or very very low (Redner 1998).

We have solved the novelty problem using networks because of its wide
application (Garlaschelli, Caldarelli, and Pietronero 2003), (Buldyrev
et al. 2010). Almost everything shows aging effect, from biological
organism to inanimate twitter ``meme''. Trend is side effect of aging;
some perform better and loose influence within few hours and some
perform for years. Therefore aging is one of the important factor that
we should taken into consideration while solving emergence phenomena.
Novelty is in one sense unexpected event or occurrence in future
temporal domain, therefore recent behavior is an important feature for
predicting rising novelties on evolving networks. Emergence can also be
considered as after product of competition and aging. Ranking index on
the basis of action of interest is one of the best way to understand
emergence and competition such as in our case action of interest is in
link formation of the node. Therefore considering all above facts we
have solved the problem of emergence or rising novelty considering;
aging and recent behavior (recent link gain) as well as total behavior
(total link gain), and tested performance of our model on ranking index
based metrics.

The rest of the paper is organized as follows. In Section 2 we have
introduced the benchmark models and also proposed our model. In section
3 we have given details about data sets and indexes that we have used
for testing our model. In section 4 we have given the results. In
Section 5 we have concluded the paper.

\section{Materials and Methods}\label{materials-and-methods}

Before introducing our two proposed models for link prediction . first
we briefly describe four existing prediction models (i.e., Node
In-degree; PageRank(PR); Popularity Based Predictor(PBP); and Temporal
Based Predictor(TBP)) which will be used to benchmark the models and
compare their performance using the evaluation metrics which will be
discussed in the last part of this Section.

\subsection{PageRank}\label{pagerank}

PageRank given by (Brin and Page 1998) was developed to rank webpages on
internet for Google search engine optimization purpose. It can be
applied in other networked architecture also where structural property
of the node plays an important role such as information diffusion,
scientific paper/author ranking etc. PageRank algorithm can be given as
follows:- If node \(n_i\) have link to node \(n_j\) there will be a
directed link between them (\(n_i->n_j\)). If node (i.e webpage) \(n_j\)
has \(S_i\) set of link to other nodes then page will distribute its
importance in \(|l_j|\)(the number of nodes in set \(|S_i|\)).
Generally, the transition matrix of the network or graph \(A\) can be
given as follows-

\[
\begin{aligned}
A_{ij}  = \left\{ {\begin{array}{*{20}c}
   {1/l_j {\  }if{\  }n_j  \in s_i }  \\
   {0{ \quad    } Otherwise}  \\
\end{array}} \right\}
\end{aligned}
\]

Since there can be nodes that do not have link to other nodes although
they are being pointed by other nodes, also known as \emph{dangling
nodes} , so new transformed matrix (S) can be given as-

\[
\begin{aligned}
{\rm S = A + N}_{{\rm cd}}
\end{aligned}
\]

Where \(N_{\rm cd}\) matrix have all the elements zero except for
\textbf{dangling nodes'} column which are \(1/N\) where \(N\) is the
number of rows or nodes in the matrix. Easy to find that those columns
are normalized that sums to one for making column stochastic matrix. Now
the PageRank of \emph{dangling nodes} will not be zero. Since random
surfer will follow the link from one page to another, suppose that
random surfer follows the PageRank (follows S) with probability
\(({\alpha})\) then there is (\(1-\alpha\)) probability that he will
choose a random page. So now PageRank matrix also known as Google matrix
\(M\) can be given as-

\[
\begin{aligned}
M = \alpha S + \frac{{(1 - \alpha )}}{n}{I_n}
\end{aligned}
\]

Where \(I_n\) is matrix of size \(n*n\), it's every element as one .
PageRank vector \(PR\) can be calculated using power method as
\(PR^k = M.PR^{^{(k - 1)} }\) it will definitely converge to a static
vector which is PageRank.

\subsection{Popularity-based
predictor(PBP)}\label{popularity-based-predictorpbp}

(ZENG et al. 2013) came up with Popularity Based Predictor(PBP). It
exploits the \emph{preferential attachment} phenomenon, which states
that popularity increases cumulatively; the rate of new link (either
item receives rating in case of Movielens, or a friend like or comments
in case of Facebook wall post activity) formation for any node is
proportional to the observed number of links which node has received in
past. If an item is popular at time \({\rm t}\), then it will probably
be popular due to the condition that current degree of an item
\({\rm k}_o {\rm (t} {\rm )}\) is a good predictor of its future
popularity. (ZENG et al. 2013) proposes to calculate the prediction
score of an item at time \(t\) can be given as follows-

\begin{equation}
{\rm s}_o  {\rm (t} {\rm ,T}_{\rm p} {\rm ) = k}_o  {\rm (t} ) - \lambda {\rm k}_o  {\rm (t} {\rm - T}_{\rm P} {\rm )}
\end{equation}

Where \(k_o (t)\) is the rating/links received up to time \(t\).
\(\lambda \in {\rm [0, 1] }\) , note that \(\lambda = 0{\rm }\) gives
the total popularity and for \(\lambda = 1{\rm }\) it gives recent
popularity. Through out the manuscript by popularity we mean number of
ratings or links received by item or node.

\subsection{Temporal Base Predictor
(TBP)}\label{temporal-base-predictor-tbp}

This model (Zhou, Zeng, and Wang 2015) considers decay effect only while
collecting score of node for future probability of getting links.

\begin{equation}
s_o (t) = \sum\limits_u {A_{uo} (t)\exp (d(T_{uo}  - t))}
\end{equation}

where \(s_o (t)\) is prediction score for node \(o\) at time \(t\),
\(A_{uo}(t)\) is the user object adjacency matrix, and \(T_{uo}\) is the
time when object or node received link and \(d\) is the decay rate.

\subsection{Proposed model 1: Recent Behaviour Dominant
Model(RBDM)}\label{proposed-model-1-recent-behaviour-dominant-modelrbdm}

In many systems such as human, herding or recent behavior followers are
very common. Although it is well known that people go for popular items
or follow rich-gets-richer behavior, but to understand emergence, recent
behavior is one of the important factor. People follow the strength of
the node also, with recent behaviours. We also know ``congestion feeds
itself'' phenomena, meaning having more recent attractions(e.g recent
new links) may lead to gaining more recent attractions. To quantify all
these phenomena we came up with a model that gives weight to recency of
link formation. If there are few attentions to node (e.g links) then
will give weight to total popularity gain. Since we are modeling a
system in which humans are core of its action for making link either
user item bipartite network, Facebook friend ship network or paper
citation network. And if people follow recent behavior in making links
then recent degree gain must be a good predictor for future degree gain.
In the same way if they follow total degree or popularity of the node
then total degree would be a good predictor. But in reality people
follow both therefore, we supppose people follow recent behavior with
probability \(\alpha\), and with \(1-\alpha\) they follow total degree
or popularity of the items.

\begin{equation}
P_o (t,T_P ) = (\alpha _o \frac{{\Delta k_o (t,T_P )}}{{\sum\limits_u {\Delta k_o (t,T_P )} }} + (1 - \alpha _o )\frac{{k_o (t)}}{{\sum\limits_u {k_o (t)} }})
\end{equation}

Where \(\Delta k_o (t,T_P ) = k_o (t) - k_o (t - T_p )\),\(k_o (t)\)
reflect the link gain upto time \(t\). \(\alpha _o\) is dominance
factor. It depends on system to system ho much recent behavior dominated
system is, weak or strong. Such as for ``bursty'' behavior strong
dominance should work. In our case \(\alpha _o\) is from Cumulative
Distribution of
\(\frac{{\Delta k_o (t,T_P )}}{{\sum\limits_u {\Delta k_o (t,T_P )} }}\).
It means \(\alpha _o\) is biased towards recent behaviours, the more
recent behaviour, the higher \(\alpha _o\) will be. IF the recent
behaviour probability
(\(\frac{{\Delta k_o (t,T_P )}}{{\sum\limits_u {\Delta k_o (t,T_P )} }}\)),
is low then the model score will depends on total link gain. Which is
why we call this model ``Recent Behavior Dominant Model(RBDM)''.

\subsection{Proposed model 2: Recent Behaviour Non-Dominant Model(RBNDM)
for
Emergence}\label{proposed-model-2-recent-behaviour-non-dominant-modelrbndm-for-emergence}

In this model we think recent behavior is effect of long term past
behavior. If a node is getting more link recently it is because it has
already gained link in past. In other words people follow recent
behaviours but also at the same time they concern about past links. In
that system we need to give weightage to both behaviours. So to model
this we think if a node is active in recent time and also it was popular
then its probability of gaining link in future is high.

\begin{equation}
P_o (t,T_P ) = ((1 - \alpha _o )\frac{{\Delta k_o (t,T_P )}}{{\sum\limits_u {\Delta k_o (t,T_P )} }} + \alpha _o \frac{{k_o (t)}}{{\sum\limits_u {k_o (t)} }})
\end{equation}

Since we are considering evolving systems, we should consider the time
as an important factor. Therefore to calculate total score of node, i.e
\(k_o (t)\) , we used aging effect similar to TBP model described above.
\(\alpha _o\) is from cumulative distribution of
\(\frac{{\Delta k_o (t,T_P )}}{{\sum\limits_u {\Delta k_o (t,T_P )} }}\).

\section{Data and Metrics}\label{data-and-metrics}

To test the performance and robustness of our model we have considered
different data sets and evaluation metrics.

\subsection{Evalutation metrics}\label{evalutation-metrics}

Four evaluation metrics are adopted to measure the accuracy of the
proposed model including
\emph{precision}\((P_n)\),\emph{novelty}\((Q_n)\) and \emph{Area Under
Recieving Operating Characteristic}(\(AUC\)) and rank correlation
Kendall's Tau(\(\tau\)).

\begin{itemize}
\item
  \emph{Precision} is defined as the fraction of objects that are
  predicted also lie in the top \(N\) object of true ranking (Herlocker
  et al. 2004). \[\begin{aligned}
   {p_n  = \frac{{D_n }}{n}}
  \end{aligned}\] Where \(D_n\) is the number of common objects between
  predicted and real ranking. \(n\) is the size of list to be ranked.
  its value ranges in {[}0,1{]}, higher value of \((P_n)\) is better.
\item
  \emph{Novelty(\(Q_n\))} is a metric to measure the ability of a
  predictor to rank the items in top \(n\) position that was not in top
  \(n\) position in past (ZENG et al. 2013). These are the rising
  novelties in the system. If we denote the predicted object as
  (\(P_po\)) and potential true object as \(P_ro\), then the novelty is
  given by- \[\begin{aligned}
  Q_n  = P_{po} /P_{ro}
  \end{aligned}\]
\item
  \emph{AUC} measures the relative position of the predicted item and
  true ranked items. Suppose predicted item list is (\(L_pn\)) and real
  item list is (\(L_rn\)). if \(s_op \in L_{pn}\) and
  \(s_rp \in L_{rn}\) is score of object in predicted then \emph{AUC} is
  given by-
\end{itemize}

\[\begin{aligned}
AUC = \frac{{\sum\limits_{op \in L_{pn} } {\sum\limits_{rp \in L_{rn} } {I(s_{pn} ,s_{rn} )} } }}{{\left| {L_{pn} } \right|\left| {L_{rn} } \right|}}
\end{aligned}\]

where, \[\begin{aligned}
I(s_{pn} ,s_{rn} ) = \left\{
   {\begin{array}{*{20}c}
   {0 \Leftarrow s_{pn}  < s_{rn} }  \\
   {0.5 \Leftarrow s_{pn} = s_{rn} }  \\
   {1 \Leftarrow s_{pn}  > s_{rn} }  \\
\end{array}} \right.
\end{aligned}\]

\begin{itemize}
\tightlist
\item
  \emph{Kendal's Tau(\(\tau\))} measures the correlation between
  predicted and actual ratings. It varies between \(-1\) and \(+1\).
  \(\tau =1\) when predicted and actual are identical,\(\tau =0\) when
  both ranking is independent and \(\tau =-1\) shows they perfectly
  disagree. It can be given as-
\end{itemize}

\[
\begin{aligned}
{\tau}= \frac{{C - D}}{{C + D}}
\end{aligned}
\]

where \(C\) is the number of concordant pairs and \(D\) is the number of
discordant pairs.

\subsection{Data}\label{data}

To test the predictors accuracy we have used different data sets
including MovieLens, Netflix, Facebook wall post and Arxiv paper
citation. MovieLens and Netflix data sets contains movie ratings,
Facebook data set contains users' wall post relationships and arxiv
citation data set contains paper citation information from arxiv
database of High Energy Physics
\href{http://snap.stanford.edu/data/cit-HepPh.html}{field} (Gehrke,
Ginsparg, and Kleinberg 2003). Movielens, Netflix and Facebook data is
same as reported in paper (Zhou, Zeng, and Wang 2015). For Arxiv high
energy physics citation data contains papers that were uploaded by
scientists. It contains \(34,546\) papers from January \(1993\) to April
\(2003\). There is an edge if paper \(i\) has cited paper \(j\). We have
converted the time into number of months. We have considered the paper
\(j\) has received link from paper \(i\) at the time paper \(i\) was
submitted to arxiv server. For the three data sets Facebook, Movielens
and Netflix the time is considered in days while for arxiv citation data
set the time is in months since citation process is slower than the rest
of the cases. The data description are as follows-

\begin{itemize}
\item
  \textbf{Netflix} data contains \(4960\) users,\(16599\) movies and
  \(1249058\) links, data was collected during(1st Jan \(2000\)
  --\(31st\) Dec 2005).
\item
  \textbf{MovieLens} data set contains \(7533\) movies, \(864581\) links
  and \(5000\) users and data was collected during(\(1st\) Jan \(2002\)
  --\(1st\) Jan 2005).
\item
  \textbf{Facebook} data contains \(40981\) set of users and their
  \(38143\) wall post activity and \(855542\) links, during period of
  (14 Sep \(2004\)--\(22nd\) Jan \(2009\)). If user has posted on a wall
  there will be a link between the user and the wall, self influenced is
  removed by removing the link between user and its own wall post.
\item
  \textbf{Arxiv-HePh} data set contains \(30500\) number of papers and
  \(347185\) edges from January \(1993\) to April \(2003\).
\end{itemize}

\fbox{\includegraphics[width=\textwidth]{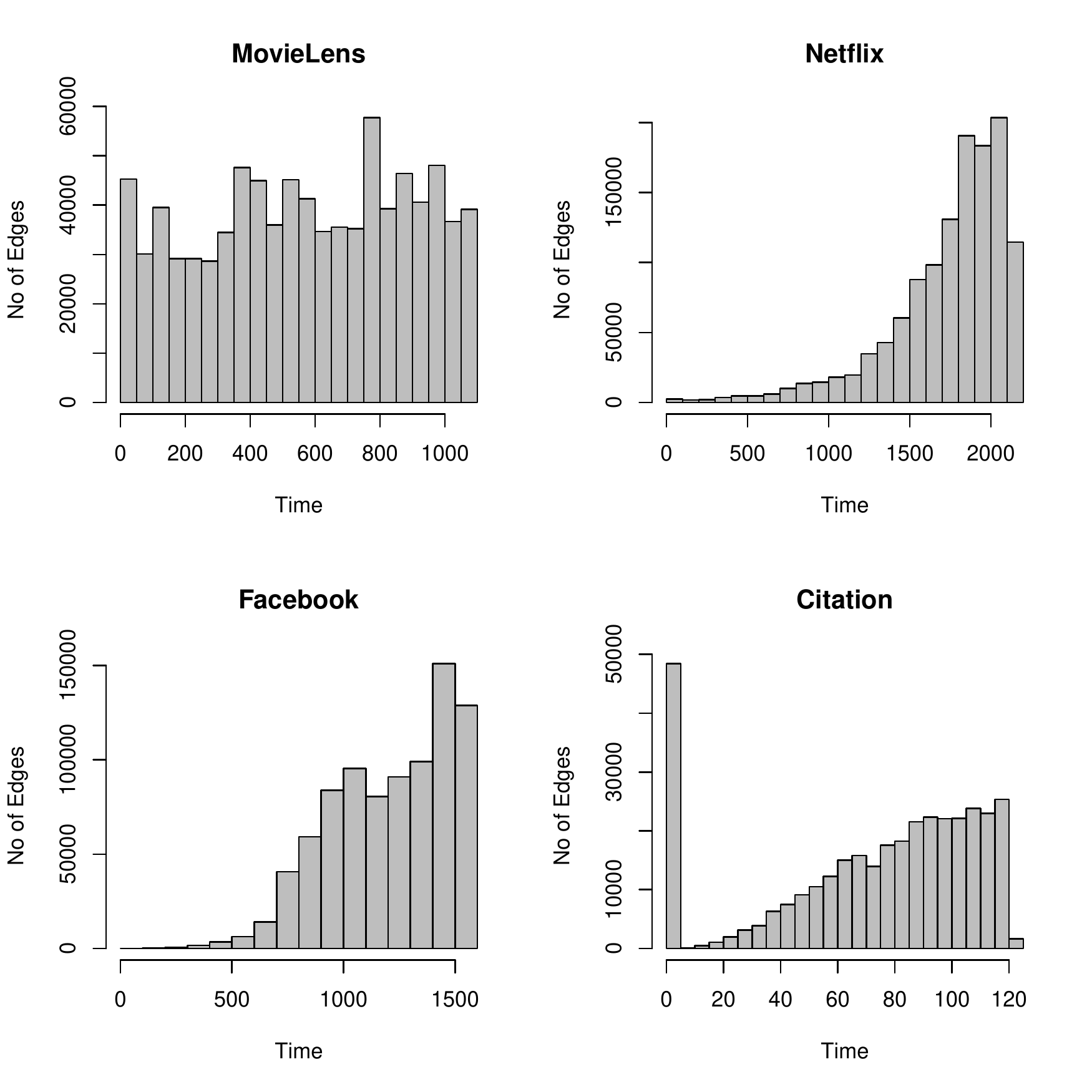}}\\

\begin{quote}
Figure 1: The above figure shows temporal occurrence of event or link
formation per unit of time, i.e for Movielens, Netflix, Facebook it is
in days and for Citation dataset it is in Months.
\end{quote}

\section{Analysis and Results}\label{analysis-and-results}

The link prediction models often require to consider the creation (and
decay) of the links in a network over time. In other words, observing
the link formation behaviour during a given time (e.g., up to the models
will predict (and rank) nodes' link gain during a period of time window
in future (\textgreater{}t). In order to compare different prediction
models, first we have calculated the values for the evaluation metrics
(Pn, Qn, AUC, tua), as shown in Figure 3, considering top 50, 100 and
200 items for all the four datasets.

\subsection{Results on synthetic data}\label{results-on-synthetic-data}

To prove our models' theoratical relevance in {[}Figure 2{]} we have
created data for recent degree gain and total degree gain. To achieve
this we have sampled data from population of size n*n, where n is total
system size. We have randomly sampled with replacement, recent degree
gain and total degree gain from population of \(1000000\). In first
column we have plotted the distribution of recent degree gain and in
second column, distribution of \(\alpha\) in our model. In third column
we have plotted the correlation rank \(\tau\) with recent degree gain
and total degree gain and with our models. It can be seen the rank
correlation for recent degree gain and Recent Behaviour Dominant
Model(Recent:RBDM) is \(1\). It proves our hypothesis ``Recent Behaviour
Dominance'' effect. In the same way we have plotted the rank correlation
with recent degree gain and RBNDM. We have repated the procedure with
total degree gain also.

\fbox{\includegraphics[width=\textwidth]{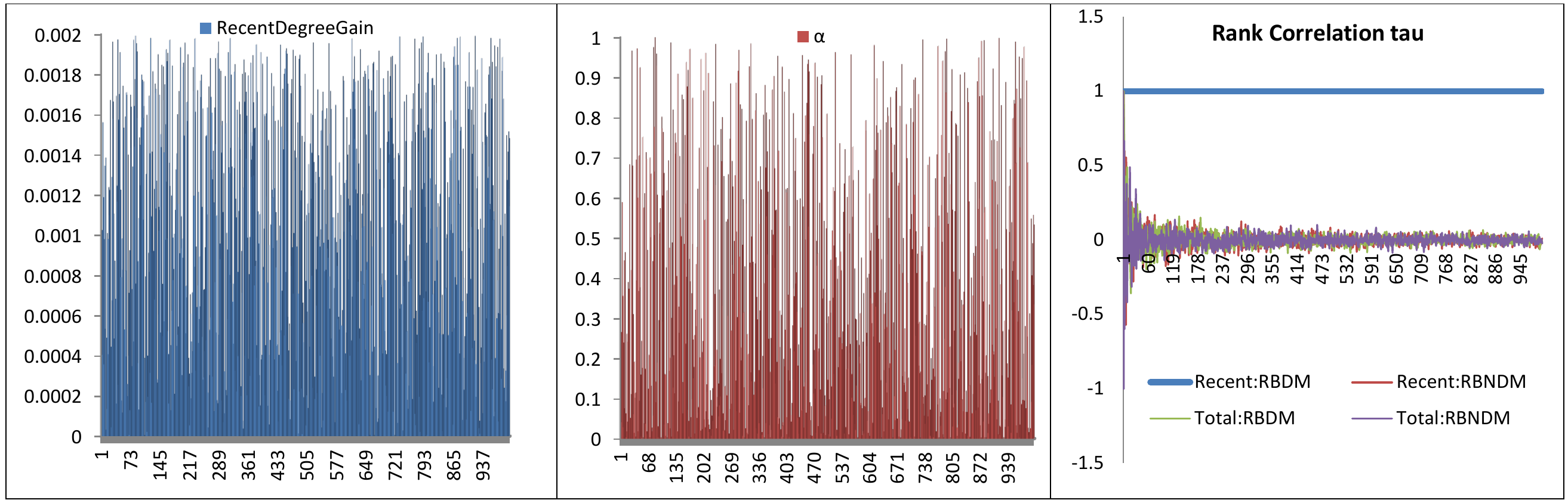}}

\begin{quote}
Figure 2: We have created random data for recent degree gain and total
degree gain and then calculated our models correlation for varying
system size. We have tested upto 1000 size. First figure from left is
probablity distribution of recent degreegain, the middle figure is
distribution of \(\alpha\) in our model and the right most is rank
correlation \(\tau\). The x axis is the size of the system. Y-axis is
the correlation value which can be between -1 to 1. Recent:RBDM means
correlation between recent degree gain and our probposed model RBDM. In
same order Total:RBNDM is the rank correlation between our proposed
model RBNDM and total degree gain, and so on.
\end{quote}

\subsection{Results on Real Data}\label{results-on-real-data}

To evaluate the performance of our models we have selected 10 random
\(t\) for each data sets. Selection of t is considered in such a way
that predictor have enough history information so we left one-third of
the data from start and one-third from last. From middle one-third part
we have randomly selected time and then calculated score. Since
predictors are based on nodes' history, we have selected only those node
that have received at least one link before time \(t\). Therefore the
new nodes after randomely selected time \(t\) will be discarded in our
analysis.

\subsection{Accuracy comparison}\label{accuracy-comparison}

For comparing our proposed predictors with benchmark predictors we have
considered past time window (\(T_P\)) and future time window (\(T_F\))
as \(30\) days and for arxiv citation data set the time is 40 months for
past and future time window both. For comparison we have selected the
top n ranked items from predicted list and compare them against the real
items for both the predictors. For pageRank teleportation parameter we
have considered \(0.90\).

While comparing our predictor we have considered the past and future
time windows as 30 days in case of Movielense, Netflix and Facebook
respectively. Paper citation case is different, its evolution takes time
so instead of day we have considered no of months. Thus in this case
past and future time windows are \(40\) months. We did this to make sure
we have enough random months so that we can take average of 10 without
any bias. Because after cleaning the data all we have 121 months start
from 0 month. \(40\) months is very short period for paper citation,
generally decay effect cant be seen in this short period of time.

\fbox{\includegraphics[width=\textwidth]{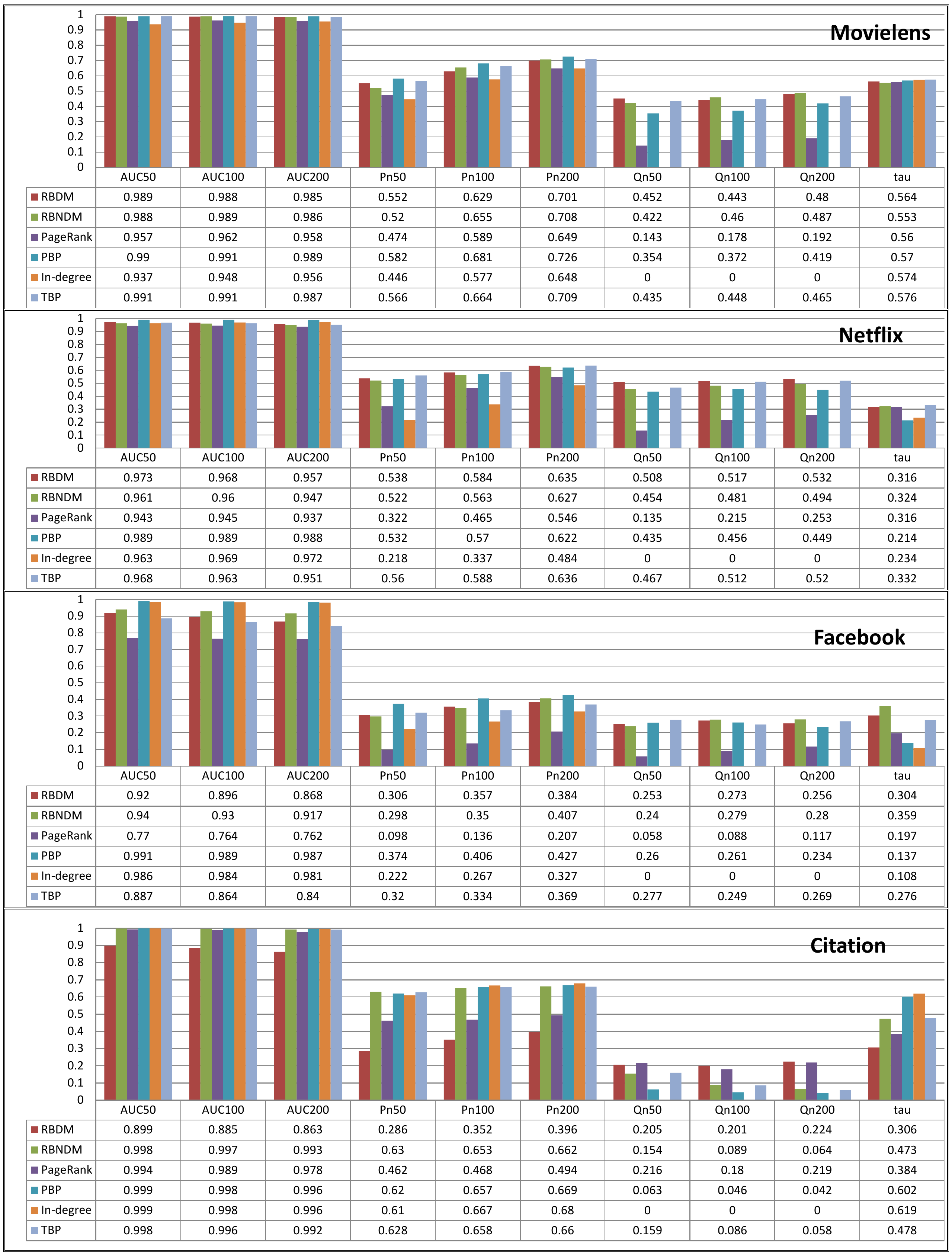}}\\

\begin{quote}
Figure 3: The performance comparison between proposed methods and
benchmark methods for MovieLens, Netflix, Facebook and citation data set
for top 50,100 and 200 items. The metrics used are; AUC, Pn , Qn and
tau, higher the better all the four metrics gives result between 0 and 1
except tau which gives between -1 to 1.
\end{quote}

We have compared our results with the benchmark methods considering top
50,100 and 200 items in list. For comparing we have considered past item
window (\(T_P\)) as 30 days and we have tested the predictor for the
same future time length \(T_F = 30\) days. For arxiv citation network we
have considered \(T_P, T_F = 40\) months. {[}Figure 3{]} shows the
comparative performance of proposed method over benchmark method. This
result is the best value achieved by predictors when \(T_F\) and \(T_P\)
are same as \(30\) days. In case of citation data set the \(T_F\) and
\(T_P\) are 40 months. The parameter value in case of TBP and in our
case is as follows: Movielens,Netflix and \(\gamma=0.06\) for Facebook
and Citation it is \(0.03\). In case of PBP \(\lambda=0.98\). The
pageRank parameter is \(0.9\) for all the datasets except Facebook in
this case \(0.6\) gives good precision. From the {[}Figure 3{]} have
better performance with benchmark models. Our proposed model not
outperforms in all the cases but there atleast one case in which it
always out performs with respect to benchmark models.

\subsection{\texorpdfstring{Perfomance for fixed past (\(T_P\)) and
varying future (T\_F) time
lenght}{Perfomance for fixed past (T\_P) and varying future (T\_F) time lenght}}\label{perfomance-for-fixed-past-tux5fp-and-varying-future-tux5ff-time-lenght}

In the next phase of our analysis, for further comparison of the
performance of different predicting models, we also investigate the
effects of the length of past and future time windows, on the
performance of models measured through Pn, Qn, AUC, and tau. .

\fbox{\includegraphics[width=\textwidth]{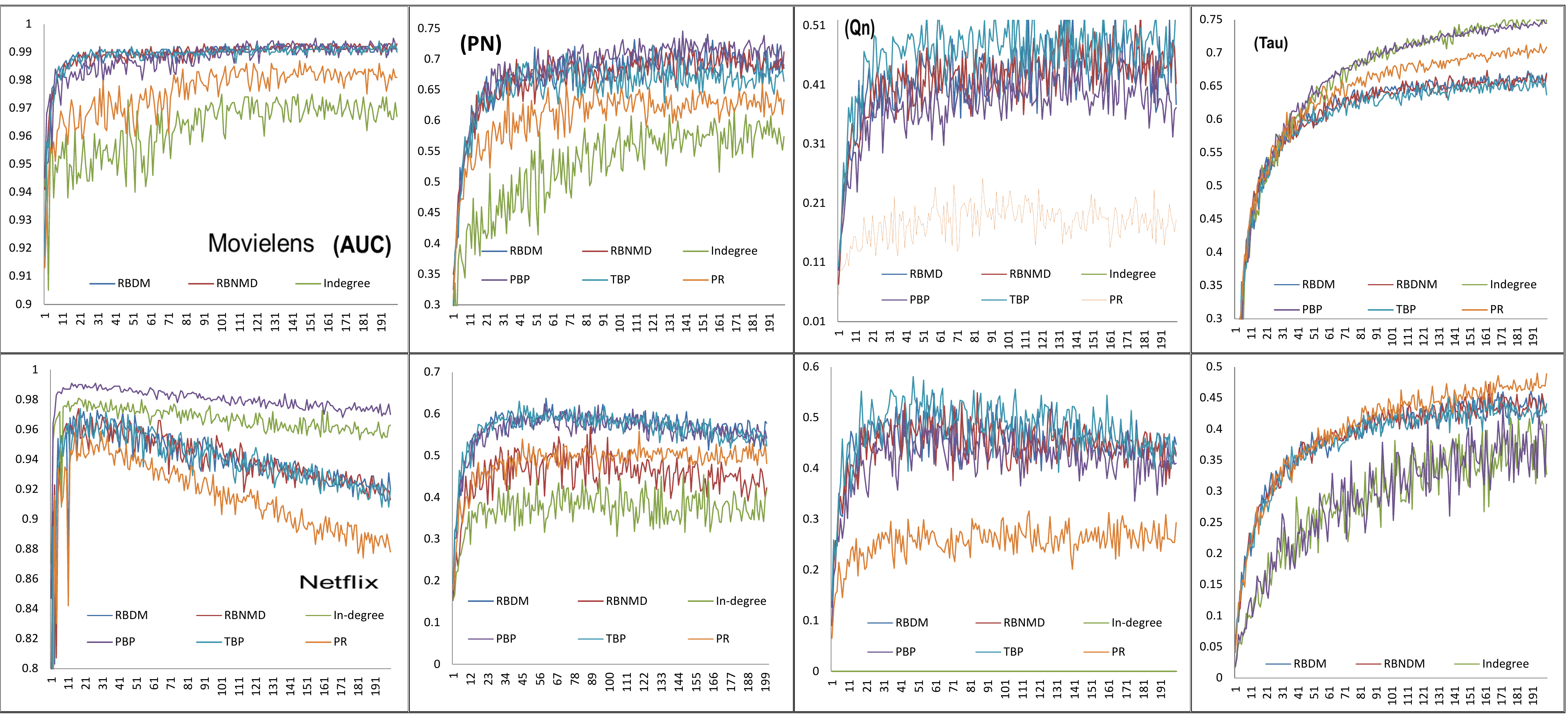}}\\
\fbox{\includegraphics[width=\textwidth]{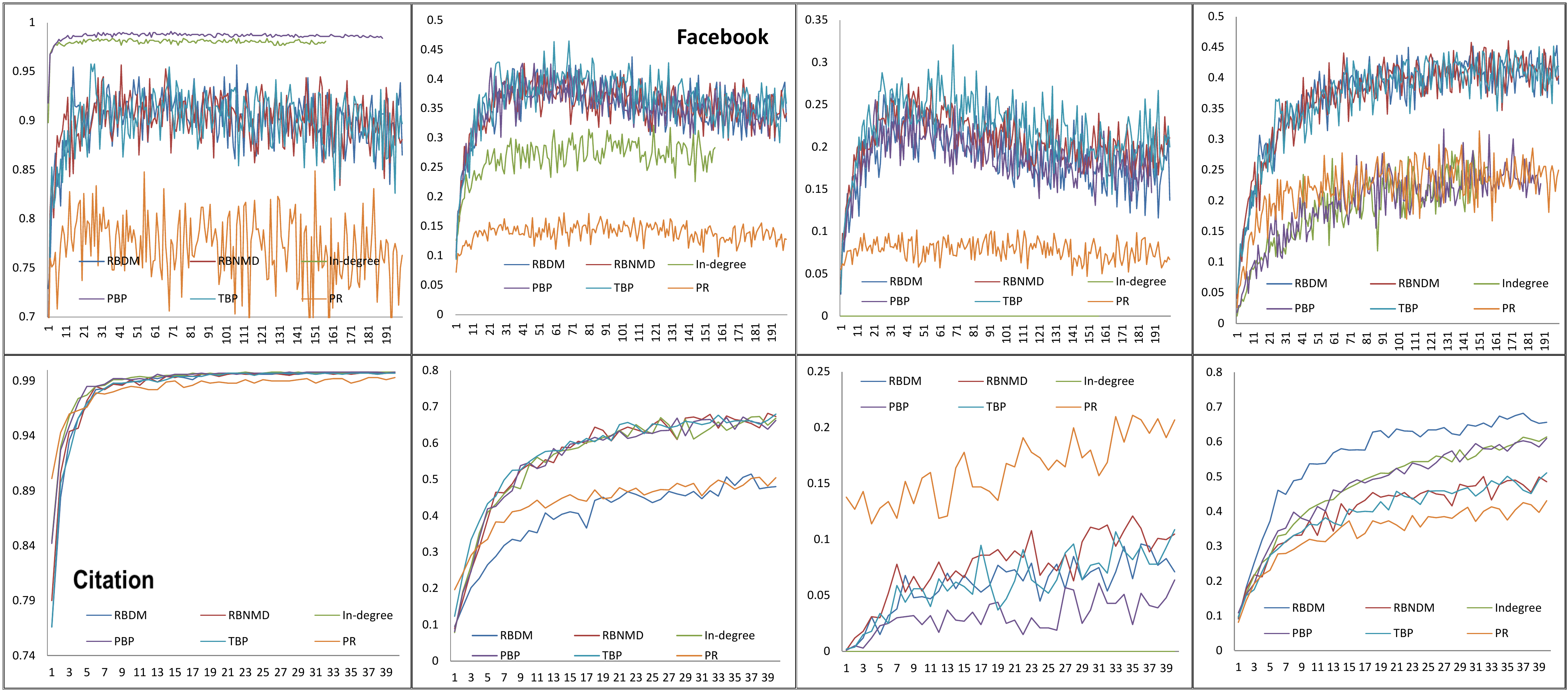}}

\begin{quote}
Figure 4: Performance of the predictor for diferent values on future
time window \(T_F\). The varticle columns are from right to left for
AUC, Pn , novelty (Qn) and rank correlation (tau). This result is for
top 100 items perfromance except rank correlation tau which works on the
whole data. Y-axis is for metrics, AUC, Pn , Qn and tau results, higher
the better.
\end{quote}

In {[}Figure 4{]} we have shown the performance of our predictor against
the benchmark predictors for different values of future time window.For
TBP and PBP the parameter values are same as the TBP author has reported
in their paper. Here we have considered past time window as fixed. We
have calculated accuracy on the basis of varying future time window. In
case of Movielens long term precision prediction is not good while Qn
prediction is good. We can find our proposed model's performance is
good. From precision analysis we can find our proposed models
performance doesn't get affected by future time window \(T_F\) for all
the datasets. Rank correlation \(\tau\) also gets better as future time
window increases for all the data sets. Novelty \(Q_n\) affected by
future time window only in case of Facebook. Our proposed model may not
out perform all the predictors in all the situation but it has
theoretical significance.

\subsection{The effects of Tp and Tf on the proposed models (RBDM and
RBNDM)
performance}\label{the-effects-of-tp-and-tf-on-the-proposed-models-rbdm-and-rbndm-performance}

In the {[}Figure 5{]} we have shown the effect of recent behavior
dominance and recent behavior non dominance in making prediction. To see
the performance we have selected top 100 items. ML-DOM implies, result
for top 100 items considering AUC index for Movielense (ML) Data set and
considering RBDM (DM-Dominant Model) and so on.

\fbox{\includegraphics[width=\textwidth]{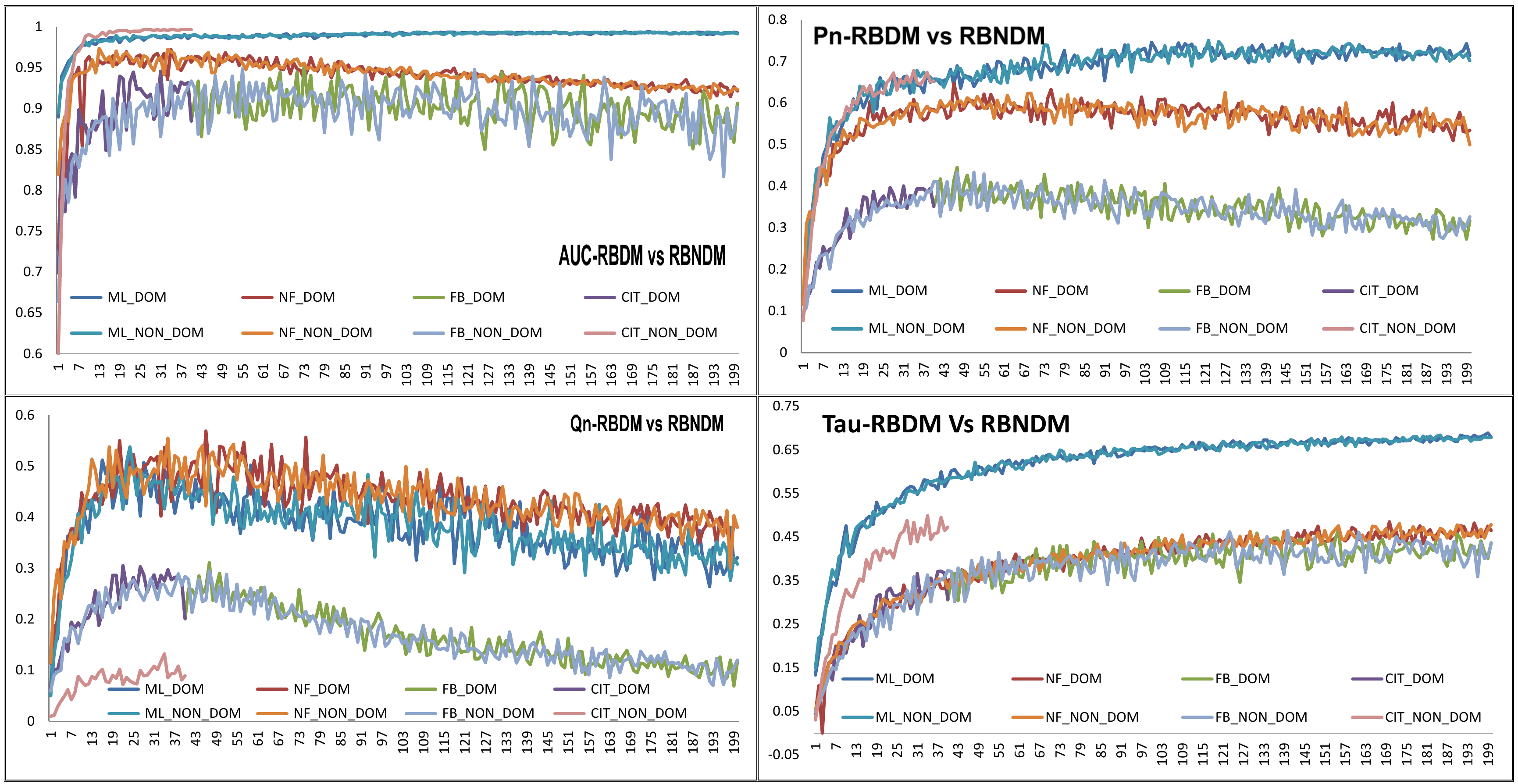}}

\begin{quote}
Figure 5: Performance of RBDM Vs RBNDM, considering past time window
(\(T_P\)) and future time window (\(T_F\)) same upto 200 days. In case
of Citation the number time is upto 40 months. Abbreviations: ML-
Movielens, NF-Netflix, FB-Facebook and CIT- Citation dataset, DOM is for
RBDM model and NON-DOM is for RBNDM. The y-axis shows the metrics value,
higher the better all the three metrics; Precision (Pn), Novelty(Qn) and
AUC gives result between 0 and 1, rank correlation tau (\(\tau\)) can
give values between -1 to 1. The x-axis shows the time duration for past
and future time window.
\end{quote}

In \(AUC\) analysis we have found that in case of Movielens short term
and long term prediction has no effect. Its performance doesn't get
affected by shorter and longer time window much. Specially in case of
Movielens. In case of Netflix both the predictors has almost similar
performance but after 100 days a slight deterioration in performance
found. Similar effect have found for Facebook also. In case of Citation
data set the time window is in terms of months, so up to 40 months its
performance increases. In case of Movielens precision (\(P_n\)) analysis
the predictor have similar performance and its accuracy are also not
getting affected. Past and future time window have negligible effect. In
\(Q_n\) or novelty prediction analysis we have found that Netflix data
has better performance over others, while both the predictor has similar
performance. The novelty prediction much affected by past and future
time window as compare to AUC and Precision. Only for predicting
citation data the both predictors performance varies much in which
predictor RBDM has better performance. In case of rank correlation both
predictor have similar performance, only in case of citation RBNDM
performs better.

\section{Conclusions}\label{conclusions}

In this manuscript we came up with two models to make prediction of node
on online social media or evolving networks specially considering its
temporal behavior. The first model (Recent Behavior Dominant
Model(RBDM)) we have considered that people go for popular as well as
follow recent behaviours but recent behavior dominates especially in
case when node has recieved more links in recent past. It suppose that
random walker walks on a network and can jump to any node, the
probablity of jumping to other nodes is based on which node is recently
more active or receiving more link. The more recently node has received
link the more it will dominate over total degree for future link gain.
The second model (Recent Behavior Non-Dominant Model(RBNDM)) in which we
think the node will gain more links if it is recently active and also
have gained link in past. We have compared our results with state of the
art models i.e popularity based predictor, pageRank , In-degree and TBP.
To test the robustness of our model we have tested our models on
different data sets which has different temporal distribution. we have
found our proposed model performs better. Our models not always out
performs the benchmarks models but in different cases on different data
sets such as RBNDM doesn't perform better than RBDM in other case but in
case of citation data set it performs better in precision, novelty and
rank correlation. Although with TBP some times it lacks. The merits of
our model is it helps identifying novel entries without any significant
cost of predicting already popular items. In this paper we have
considered the recent behavior as the basis of identifying rising
novelties with other combinations such as aging, dominant recent
behavior as well as non dominant recent behavior. In all of these with
respect to our data sets prediction accuracy is good. We have considered
only temporal effects of the node's attracting new link. We have found
it one of the important feature for making prediction and considering
other complex feature might increase accuracy but it increases
computational complexity. So we can say our model computational resource
efficient also. In our current work we have not considered any threshold
when recent behavior will start dominating over total behavior in future
work one can consider the threshold.

\subsection{Acknowledgement}\label{acknowledgement}

This work was supported by the National Natural Science Foundation of
China (Grant Nos. 61370150 and 61433014).

\section*{References}\label{references}
\addcontentsline{toc}{section}{References}

\hyperdef{}{ref-greycite57259}{\label{ref-greycite57259}}
Abbasi, Alireza. 2016. ``A Longitudinal Analysis of Link Formation on
Collaboration Networks.'' \emph{Journal of Informetrics}.
\url{http://www.sciencedirect.com/science/article/pii/S1751157716300815}.
\url{http://dx.doi.org/10.1016/j.joi.2016.05.001}.

\hyperdef{}{ref-Axtellux5f2001}{\label{ref-Axtellux5f2001}}
Axtell, R. L. 2001. ``Zipf Distribution of U.S. Firm Sizes.''
\emph{Science} 293 (5536). American Association for the Advancement of
Science (AAAS): 1818--20.
\href{http://doi.org/10.1126/science.1062081}{doi:10.1126/science.1062081}.

\hyperdef{}{ref-Bouchaudux5f2001}{\label{ref-Bouchaudux5f2001}}
Bouchaud, Jean-Philippe, Andrew Matacz, and Marc Potters. 2001.
``Leverage Effect in Financial Markets: The Retarded Volatility Model.''
\emph{Phys. Rev. Lett.} 87 (22). American Physical Society (APS).
\href{http://doi.org/10.1103/physrevlett.87.228701}{doi:10.1103/physrevlett.87.228701}.

\hyperdef{}{ref-Brinux5f1998}{\label{ref-Brinux5f1998}}
Brin, Sergey, and Lawrence Page. 1998. ``The Anatomy of a Large-Scale
Hypertextual Web Search Engine.'' \emph{Computer Networks and ISDN
Systems} 30 (1-7). Elsevier BV: 107--17.
\href{http://doi.org/10.1016/s0169-7552(98)00110-x}{doi:10.1016/s0169-7552(98)00110-x}.

\hyperdef{}{ref-Buldyrevux5f2010}{\label{ref-Buldyrevux5f2010}}
Buldyrev, Sergey V., Roni Parshani, Gerald Paul, H. Eugene Stanley, and
Shlomo Havlin. 2010. ``Catastrophic Cascade of Failures in
Interdependent Networks.'' \emph{Nature} 464 (7291). Springer Nature:
1025--28.
\href{http://doi.org/10.1038/nature08932}{doi:10.1038/nature08932}.

\hyperdef{}{ref-DBLP:confux2fwwwux2fChengAKL16}{\label{ref-DBLP:confux2fwwwux2fChengAKL16}}
Cheng, Justin, Lada A. Adamic, Jon M. Kleinberg, and Jure Leskovec.
2016. ``Do Cascades Recur?'' In \emph{Proceedings of the 25th
International Conference on World Wide Web, WWW 2016, Montreal, Canada,
April 11 - 15, 2016}, edited by Jacqueline Bourdeau, Jim Hendler, Roger
Nkambou, Ian Horrocks, and Ben Y. Zhao, 671--81. ACM.
\href{http://doi.org/10.1145/2872427.2882993}{doi:10.1145/2872427.2882993}.

\hyperdef{}{ref-Eux5fux5f1925}{\label{ref-Eux5fux5f1925}}
E., F. Y., and G. Udny Yule. 1925. ``A Mathematical Theory of Evolution
Based on the Conclusions of Dr. J. c. Willis, F.R.S.'' \emph{Journal of
the Royal Statistical Society} 88 (3). JSTOR: 433.
\href{http://doi.org/10.2307/2341419}{doi:10.2307/2341419}.

\hyperdef{}{ref-Eislerux5f2008}{\label{ref-Eislerux5f2008}}
Eisler, Zolt\a'an, Imre Bartos, and J\a'anos Kert\a'esz. 2008.
``Fluctuation Scaling in Complex Systems: Taylors Law and Beyond1.''
\emph{Advances in Physics} 57 (1). Informa UK Limited: 89--142.
\href{http://doi.org/10.1080/00018730801893043}{doi:10.1080/00018730801893043}.

\hyperdef{}{ref-Garlaschelliux5f2003}{\label{ref-Garlaschelliux5f2003}}
Garlaschelli, Diego, Guido Caldarelli, and Luciano Pietronero. 2003.
``Universal Scaling Relations in Food Webs.'' \emph{Nature} 423 (6936).
Springer Nature: 165--68.
\href{http://doi.org/10.1038/nature01604}{doi:10.1038/nature01604}.

\hyperdef{}{ref-Gehrkeux5f2003}{\label{ref-Gehrkeux5f2003}}
Gehrke, Johannes, Paul Ginsparg, and Jon Kleinberg. 2003. ``Overview of
the 2003 KDD Cup.'' \emph{SIGKDD Explor. Newsl.} 5 (2). Association for
Computing Machinery (ACM): 149.
\href{http://doi.org/10.1145/980972.980992}{doi:10.1145/980972.980992}.

\hyperdef{}{ref-Ginsbergux5f2008}{\label{ref-Ginsbergux5f2008}}
Ginsberg, Jeremy, Matthew H. Mohebbi, Rajan S. Patel, Lynnette Brammer,
Mark S. Smolinski, and Larry Brilliant. 2008. ``Detecting Influenza
Epidemics Using Search Engine Query Data.'' \emph{Nature} 457 (7232).
Springer Nature: 1012--14.
\href{http://doi.org/10.1038/nature07634}{doi:10.1038/nature07634}.

\hyperdef{}{ref-gleeson2014simple}{\label{ref-gleeson2014simple}}
Gleeson, James P, Davide Cellai, Jukka-Pekka Onnela, Mason A Porter, and
Felix Reed-Tsochas. 2014. ``A Simple Generative Model of Collective
Online Behavior.'' \emph{Proceedings of the National Academy of
Sciences} 111 (29). National Acad Sciences: 10411--15.
\href{http://doi.org/10.1073/pnas.1313895111}{doi:10.1073/pnas.1313895111}.

\hyperdef{}{ref-Haldaneux5f2011}{\label{ref-Haldaneux5f2011}}
Haldane, Andrew G., and Robert M. May. 2011. ``Systemic Risk in Banking
Ecosystems.'' \emph{Nature} 469 (7330). Springer Nature: 351--55.
\href{http://doi.org/10.1038/nature09659}{doi:10.1038/nature09659}.

\hyperdef{}{ref-Herlockerux5f2004}{\label{ref-Herlockerux5f2004}}
Herlocker, Jonathan L., Joseph A. Konstan, Loren G. Terveen, and John T.
Riedl. 2004. ``Evaluating Collaborative Filtering Recommender Systems.''
\emph{ACM Transactions on Information Systems} 22 (1). Association for
Computing Machinery (ACM): 5--53.
\href{http://doi.org/10.1145/963770.963772}{doi:10.1145/963770.963772}.

\hyperdef{}{ref-Hommesux5f2002}{\label{ref-Hommesux5f2002}}
Hommes, C. H. 2002. ``Modeling the Stylized Facts in Finance Through
Simple Nonlinear Adaptive Systems.'' \emph{Proceedings of the National
Academy of Sciences} 99 (Supplement 3). Proceedings of the National
Academy of Sciences: 7221--28.
\href{http://doi.org/10.1073/pnas.082080399}{doi:10.1073/pnas.082080399}.

\hyperdef{}{ref-Hoppeux5f1984}{\label{ref-Hoppeux5f1984}}
Hoppe, Fred M. 1984. ``Polya-Like Urns and the Ewens Sampling Formula.''
\emph{Journal of Mathematical Biology} 20 (1). Springer Nature: 91--94.
\href{http://doi.org/10.1007/bf00275863}{doi:10.1007/bf00275863}.

\hyperdef{}{ref-Keux5f2015}{\label{ref-Keux5f2015}}
Ke, Qing, Emilio Ferrara, Filippo Radicchi, and Alessandro Flammini.
2015. ``Defining and Identifying Sleeping Beauties in Science.''
\emph{Proceedings of the National Academy of Sciences} 112 (24).
Proceedings of the National Academy of Sciences: 7426--31.
\href{http://doi.org/10.1073/pnas.1424329112}{doi:10.1073/pnas.1424329112}.

\hyperdef{}{ref-Kingux5f2011}{\label{ref-Kingux5f2011}}
King, G. 2011. ``Ensuring the Data-Rich Future of the Social Sciences.''
\emph{Science} 331 (6018). American Association for the Advancement of
Science (AAAS): 719--21.
\href{http://doi.org/10.1126/science.1197872}{doi:10.1126/science.1197872}.

\hyperdef{}{ref-Paroloux5f2015}{\label{ref-Paroloux5f2015}}
Parolo, Pietro Della Briotta, Raj Kumar Pan, Rumi Ghosh, Bernardo A.
Huberman, Kimmo Kaski, and Santo Fortunato. 2015. ``Attention Decay in
Science.'' \emph{Journal of Informetrics} 9 (4). Elsevier BV: 734--45.
\href{http://doi.org/10.1016/j.joi.2015.07.006}{doi:10.1016/j.joi.2015.07.006}.

\hyperdef{}{ref-Percux5f2012}{\label{ref-Percux5f2012}}
Perc, M. 2012. ``Evolution of the Most Common English Words and Phrases
over the Centuries.'' \emph{Journal of The Royal Society Interface} 9
(77). The Royal Society: 3323--28.
\href{http://doi.org/10.1098/rsif.2012.0491}{doi:10.1098/rsif.2012.0491}.

\hyperdef{}{ref-Preisux5f2013}{\label{ref-Preisux5f2013}}
Preis, Tobias, Helen Susannah Moat, and H. Eugene Stanley. 2013.
``Quantifying Trading Behavior in Financial Markets Using Google
Trends.'' \emph{Sci. Rep.} 3 (April). Nature Publishing Group.
\href{http://doi.org/10.1038/srep01684}{doi:10.1038/srep01684}.

\hyperdef{}{ref-Rednerux5f1998}{\label{ref-Rednerux5f1998}}
Redner, S. 1998. ``How Popular Is Your Paper? An Empirical Study of the
Citation Distribution.'' \emph{The European Physical Journal B} 4 (2).
Springer Nature: 131--34.
\href{http://doi.org/10.1007/s100510050359}{doi:10.1007/s100510050359}.

\hyperdef{}{ref-Simonux5f1955}{\label{ref-Simonux5f1955}}
Simon, Herbert A. 1955. ``On a Class of Skew Distribution Functions.''
\emph{Biometrika} 42 (3/4). JSTOR: 425.
\href{http://doi.org/10.2307/2333389}{doi:10.2307/2333389}.

\hyperdef{}{ref-Triaux5f2014}{\label{ref-Triaux5f2014}}
Tria, F., V. Loreto, V. D. P. Servedio, and S. H. Strogatz. 2014. ``The
Dynamics of Correlated Novelties.'' \emph{Sci. Rep.} 4 (July). Nature
Publishing Group.
\href{http://doi.org/10.1038/srep05890}{doi:10.1038/srep05890}.

\hyperdef{}{ref-Vespignaniux5f2009}{\label{ref-Vespignaniux5f2009}}
Vespignani, A. 2009. ``Predicting the Behavior of Techno-Social
Systems.'' \emph{Science} 325 (5939). American Association for the
Advancement of Science (AAAS): 425--28.
\href{http://doi.org/10.1126/science.1171990}{doi:10.1126/science.1171990}.

\hyperdef{}{ref-Wangux5f2013}{\label{ref-Wangux5f2013}}
Wang, D., C. Song, and A.-L. Barabasi. 2013. ``Quantifying Long-Term
Scientific Impact.'' \emph{Science} 342 (6154). American Association for
the Advancement of Science (AAAS): 127--32.
\href{http://doi.org/10.1126/science.1237825}{doi:10.1126/science.1237825}.

\hyperdef{}{ref-Zanetteux5f2005}{\label{ref-Zanetteux5f2005}}
Zanette, Dami\a'an, and Marcelo Montemurro. 2005. ``Dynamics of Text
Generation with Realistic Zipfs Distribution.'' \emph{Journal of
Quantitative Linguistics} 12 (1). Informa UK Limited: 29--40.
\href{http://doi.org/10.1080/09296170500055293}{doi:10.1080/09296170500055293}.

\hyperdef{}{ref-ZENGux5f2013}{\label{ref-ZENGux5f2013}}
ZENG, AN, STANISLAO GUALDI, MAT\a'UŠ MEDO, and YI-CHENG ZHANG. 2013.
``Trend Prediction in Temporal Bipartite Networks: The Case of
Movielens, Netflix, and Digg.'' \emph{Advances in Complex Systems} 16
(04n05). World Scientific Pub Co Pte Lt: 1350024.
\href{http://doi.org/10.1142/s0219525913500240}{doi:10.1142/s0219525913500240}.

\hyperdef{}{ref-Zhouux5f2015}{\label{ref-Zhouux5f2015}}
Zhou, Yanbo, An Zeng, and Wei-Hong Wang. 2015. ``Temporal Effects in
Trend Prediction: Identifying the Most Popular Nodes in the Future.''
Edited by Xia Li. \emph{PLOS ONE} 10 (3). Public Library of Science
(PLoS): e0120735.
\href{http://doi.org/10.1371/journal.pone.0120735}{doi:10.1371/journal.pone.0120735}.

\end{document}